# A Digital Twin Framework for Cyber Security in Cyber-Physical Systems

Tianyu Zhao[1, *], Ernest Foo[2] and Hui Tian[2]

[1]School of Information Technology, Griffith University, Brisbane, Australia
[2]School of Information Technology, Griffith University, Brisbane, Australia
[3]School of Information Technology, Griffith University, Brisbane, Australia

*Corresponding author e-mail: Tianyu.Zhao@griffithuni.edu.au

**Abstract**. Currently, most of the research in digital twins focus on simulation and optimization. Digital twins are especially useful for critical systems. However, digital twins can also be used for safety and cyber security. The idea of this paper is motivated by the limitations of cyber security in Cyber-Physical Systems (CPSs). We introduce an efficient synchronization approach to maintain state between virtual environment and the physical environment. In this case, we can receive prompt feedback from conducting security analysis in the virtual domain. Thus, helping to enhance the cyber security of CPSs, we propose a digital twin-based framework. Based on the approach, the security of the CPSs can be protected by the digital twin system. Moreover, the proposed architecture has also been optimized to meet the security requirements and maintain less network burden for CPSs

## 1. Introduction

A Cyber- physical system (CPS) is an internet connected system that integrates computational and physical assets. CPSs can realize real-time monitoring and dynamic control in autonomous systems. In recent years, the security in CPSs has been neglected in many industries. For example, the security attack at Maroochy Water Service in 2000 [1], the Havex worm at Europe in 2014 [2] and the cyber-attack at Ukraine power in 2016 [3]. These events have noted that the cyber-attacks in CPSs may cause huge impacts for the public.

Compared to the traditional IT system, CPSs always have a large number of legacy systems [4]. In CPSs, safety is a crucial issue needed to be considered. Based on a report published by Dragos, Inc. in 2017 [5], it highlighted that 64% of vulnerability patches do not get fixed because of insecure design. Operators are unable to take another action, even when vulnerability patches have been released. Consequently, it may cause a devastating result to the whole system. While the security in CPSs has raised more attention during the past a few years, many organizations have considered cyber security as important. Moreover, industrial standards have been developed to support public safety such as IEC 62443, VDI/VDE 2182 and NIST SP 800-82[8][9].

However, a digital twin system could be a possibly solve this problem. A digital twin can be considered a virtual replica of its physical counterpart. It could run independently in a virtual environment. In this case, operators could conduct security tests and analyse security manners without taking the risk to interfere with the physical domain. On the other hand, by using relevant feedback from the virtual environment, operators could use the information to enhance the security of the physical environment.

In this paper, we propose a digital twin framework for the cyber security of CPSs. In this case, we focus on the efficient synchronization between the physical domain and the virtual domain. Only if the real-time data transmission promptly transfers between those two domains, the security of the CPSs can be enhanced by the feedback from the virtual domain. In our digital twin framework, we predefine the key states of devices in the physical domain. In this case, the transmitted data is largely decreased in the synchronization stage. Then, based on a regular time slot, we transmit the key states to the virtual domain. The virtual replica receives significant features from its physical counterparts. In order to simulate physical devices, we use the state machine concept with the digital twin system to urge the consistence in that framework.

The paper is structured as follows: Section 2 is the literature review of previous works. In the section 3, we propose a new digital twin framework which contains the state replication model, which concentrates on efficient synchronization of the digital twin network. In the section 4, we discuss the feasibility and conclude for the future work.

## 2. Previous Work

In this section, it mainly provides detailed information about existing research related with our topic. As explained in the previous section, the digital twin is a new concept that used in the CPSs for optimization. Only in the recent years, the digital twin starts to consider as a possible solution to solve security problems and challenges in the CPSs. Therefore, there is a limit number of works focus on that area. we explain each paper and figure out the possibilities of the improvements.

In the work [10], Bitton et al focused on the balance between budget and fidelity. They proposed a framework to construct a digital twin system, which is network-specific, cost-efficient, highly reliable and security test-oriented. In their framework, the digital twin system consists of two modules, which including a problem builder and a solver. In the problem builder, it collects data from the CPSs and converts those data into a rule set, which can reflect the topology of the system and the digital twin implementation constraints. In the solver, it can find an optimal solution for security problems in CPSs by using 0-1 non-linear programming. Furthermore, the authors demonstrated their approach by using a simplified ICS network. In generally, the authors were more focused on the trade-off between budget and accuracy rather than the digital twin itself. In additional, they did not show a clear explanation about the data exchange between the physical environment and the virtual environment. More importantly, the physical system used in this research is not generic. Hence, the way to construct the digital twin system cannot fit into any CPSs.

Eckhart and Ekelhart [11] provided a novel framework, which allow users to build a digital twin system based on their physical systems. In their framework, they aim to automatically generate the virtual environment from the specifications, which captured from physical systems. Therefore, security personnel can conduct safety experiment in the virtual environment without affecting the operations in live systems. Furthermore, the authors demonstrated the viability of their framework, which including the generation of the digital twin system and security analysis. The biggest challenge from this work is the limitation of data format and network protocols. In their work, some types of data, such as, floats and string, are not available in their experiments. In other words, it does not support all kinds of data in their proposed framework. Moreover, some data needed to be manually input as the automation scheme is not completely finalized.

Eckhart et al. later extended further the work. They proposed a framework that provided a security-aware environment for digital twin systems, which demonstrated how security and safety rules can be monitored in security-relevant use cases. In [12], Eckhart et al. extended the framework using a specification-based, physical device state replication approach, by passively monitoring their inputs and outputs, showing successful detections of attacks against a CPS testbed. In their framework, they first identified some specifications in the physical system. Then, they replicate those data into the virtual environment. During the replication process, they used state machine technology to ensure consistency between the physical environment and the virtual environment. However, relying on the properties of the device, inferring the devices' characteristics based on the role definition may be unviable. Another possible option, yet more complex to implement, may be to develop a code analyzer that automatically inspects all programs, which are referenced in a CPS's specification, for the purpose of stimuli identification.

To resolve the general security issues, Gehrmann and Gunnarsson [13] considered the digital twin model as an enabler to improve security in CPSs. In their framework, they initially identified design-driving security requirements for digital twin based on data sharing and control. Furthermore, they also used the state machine technology to ensure the state synchronization between the physical domain and the virtual domain. More importantly, they used time as the parameter to trigger the synchronization in the digital twin. The shortcomings in this work are that they limited the scenario for low complexity digital twin system with moderate synchronization frequencies. Moreover, they made first proof of concept of the architecture. In order to test the effect of the proposed architecture on a great diversity of platforms and production scenarios, we need to conduct more performance evaluations on a wide variety of platforms. In their work, it has proven the consistency of the proposed synchronization protocol and showed that the security of the protocol relies on the security of the underlying used secure channel. Therefore, it is necessary to do formal analysis of the security of the complete system design and all protocols.

Inspired from the related works, we also use the state machine concept with the digital twin system to capture states to reflect the physical devices in the virtual domain. Moreover, we propose a novel digital twin framework as a protector to secure CPSs. Our approach aims to synchronize key states in the digital twin system based on defined time slot. More importantly, in our framework, we provide the lightweight network mechanism for the synchronization in the digital twin systems.

## 3. Digital Twin based Framework

In this section, we introduce a detailed description of a digital twin framework. Firstly, we provide the definitions for each component in the digital twin system. Then, we summarize the threat model. Furthermore, we identify the security requirements for CPSs. Later, we discuss properties of each part in the framework. Lastly, we present the synchronization process by using state machine replication to achieve the identified requirements.

*3.1. Definitions and Notations*

In our proposed digital twin framework, we let by $d \in D$, where d means a physical twin device and D is the set of physical devices in the physical domain. Let $d' \in D'$, where d' is the digital twin and D' is the set of the digital twins in the virtual domain. We can define the key states of the physical twin as a set $S_d = \{S_{d0}, s_{d1}, s_{d2},.., s_{dm-1}\}$ and the key states in the digital twin as a set $S_{d'} = \{S_{d'0}, s_{d'1}, s_{d'2},.., s_{d'n-1}\}$, m>n. The key states are the states which have significant properties. For example, if boil water from 0 degree to 100 Celsius, the key state is 100 degrees in that process. We assign the physical twin as the main controller of the system. Hence, key states in the digital twin are smaller than the states in the physical twin. Moreover, we let $I_d = \{i_{d0}, i_{d1}, \ldots, i_{dp-1}\}$ represent a set of inputs deployed in the physical twins and $I_{d'} = \{i_{d'0}, i_{d'1}, \ldots, i_{d'q-1}\}$ represent a set of inputs to the digital twins. We assume the synchronization can be conducted at a certain period. We let t to represent a time slot. Hence, say that $s_{d,t} \in S_d$, which means the key state of a physical twin d at a time slot t. In the same way, the key state of a digital twin d' at a time slot t could represent as $s_{d',t} \in S_{d'}$. For the inputs, we use a similar approach based on the time t. in this case, we denote $i_{d,t} \in I_d$ to represent the input to the physical twin d at time t and $i_{d',t} \in I_{d'}$ means the input to the digital twin d' at time t. Finally, we have the initial state of the physical twin and digital twin respectively: $s_{d,o}$ and $s_{d',0}$. The transition function of physical twin and the digital twin could use symbol $\delta_d : S_d \times I_d \rightarrow S_d$ and $\delta_{d'} : S_{d'} \times I_{d'} \rightarrow S_{d'}$. The synchronized data needs to be transmitted in an encryption format for protection, which we use $e_d \rightarrow e_{d'}$ or $e_{d'} \rightarrow e_d$ based on the time t.

*3.2. Threat Model*

The CPSs have affected by various types of cyber-attacks, which may include the Eavesdropping, DoS Attack, Stealthy Deception Attack, Jamming Attack, Compromised-Key Attack, Man-in-the-Middle

Attack and etc [14]. Furthermore, we can narrow down those attacks into four types, which is Message Delete, Message Insert, Message Modify and Message Replay. They have different damages to the virtual domain and the physical domain. In the virtual domain, the Message Delete causes sensor data missed. The Message Modify and the Message Replay cause false sensor data received. The Message Replay causes incorrect data or state received. For the physical domain, the Message Delete causes physical devices do not react. The Message Insert and the Message Modify can cause physical devices under security attacks. In this case, the attacked physical devices do not follow orders from the systems. The Message Replay cause the physical devices repeat previous states.

*3.3. Requirements*

We have used threat model to identify the scope of the adversary for the digital twin framework. In this section, we provide a detailed explanation about objectives for our framework. At the meantime, we illustrate the importance of synchronization regarding to the security of CPSs.

Requirement 1: The state of the virtual twin must match that of the physical Twin. The synchronizations between the physical domain and the virtual domain need to be consistence all the time.

Requirement 2: The virtual twin must have sufficient data to simulate the state of the physical twin. Since the virtual domain need to protect the CPSs in the physical domain, the virtual replicas need to be simulated close to their physical counterparts.

Requirement 3: The physical twin must have sufficient data and time to respond to virtual twin information. The outcomes from the security analysis in the virtual domain are very useful. Those outcomes can be used to improve the security of the CPSs. Hence, the synchronization is the channel to connect those two domains to communicate and transfer data.

The above requirements can build a secure environment for our framework. In the virtual domain, the Message Delete and the Message Replay can be detected by the Requirement 1. The Message Insert and the Message modify can be stopped by the Requirement 1 and 2. In the physical domain, all those four attacks can be stopped by the Requirement 1 and 3.

*3.4. Overview of the Framework*

As shown in Fig.1, the overall architecture can be divided into virtual environment and physical environment. From the previous section, we understand the traditional way to do security analysis could affect the daily operations in the CPSs. Therefore, we replicate the CPS as a virtual CPS by using digital twin system. Then, we could conduct security analysis in the virtual domain and receive feedbacks from those analysis. Based on the feedbacks, we could conduct related strategies to enhance the security of CPSs. However, to achieve that goal, we need to build an efficient synchronization between physical domain and the virtual domain. Therefore, in this framework, we only focus on synchronization in the digital twin system. This paper provides an overview of the framework. More research needed for specific applications. In the following subsections, each important part of the architecture will be illustrated.

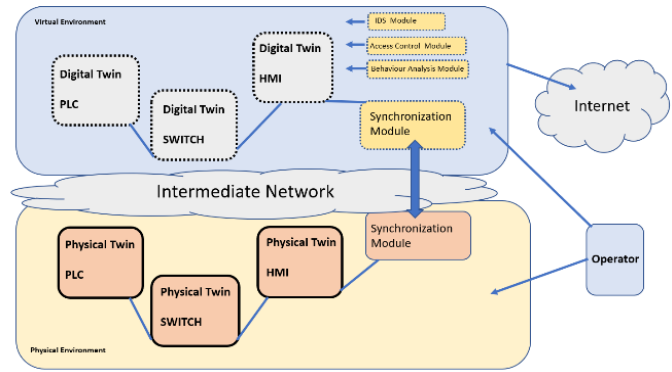

Fig. 1 Overview of the digital twin framework

*Digital twin component:* The digital twin components run independently in the virtual environment. An overview of the logical function of the digital twin component is given in Fig. 2. The main part of the digital twin is the digital simulator, which continuously simulates the states of its physical counterpart. Because the digital twin can connect with the external server, it means that there are two interconnections with the digital simulator. One is the input from the operator, who can launch a command to configure the digital simulator. Another can relate with the changes of twin states. For example, the changes of the twin states drive the digital simulator move forward. At the same time, the twin states can restore the changes once something happens in the digital simulator. All the states can be exported to the IDS module for further analysis. An anomaly-based intrusion detection can be used in this module. The log module restores all the information regarding the changes of the twin states. The purpose of this module is to replicate the virtual device corresponding to its physical counterpart. Moreover, the information is exported to the analysis module, which authenticates information with the data from the physical environment. Eventually, the synchronization module transfers all the data in virtual environment to the physical domain and vice versa.

*Physical component:* The overall framework of the logical function of the physical twin is shown in Fig. 3. It is similar to the framework of digital twin. The main part of the physical twin is the physical counterpart. This is the real device deployed in the physical domain. Similarly, it has the connections with the external operators. Besides, twin states can drive the physical counterpart to the next state. In this case, the twin state module could record all the state changes of the physical device. Due to the security concerns, these changes are recorded in the log. Furthermore, the data can be exported to analysis module to be authenticated. Finally, the synchronization module can transfer the states of the physical device to the virtual environment. At the same time, the physical domain receives the commands from the virtual environment through the synchronization module.

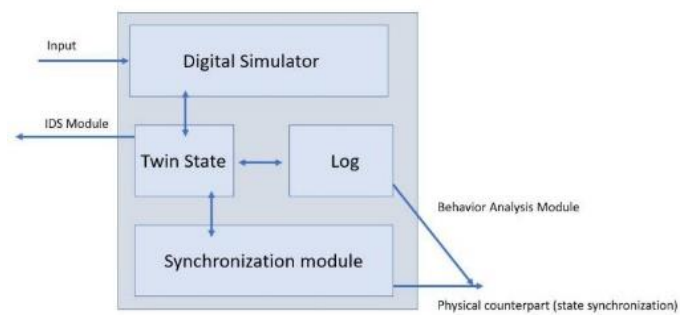

Fig. 2 The digital twin component

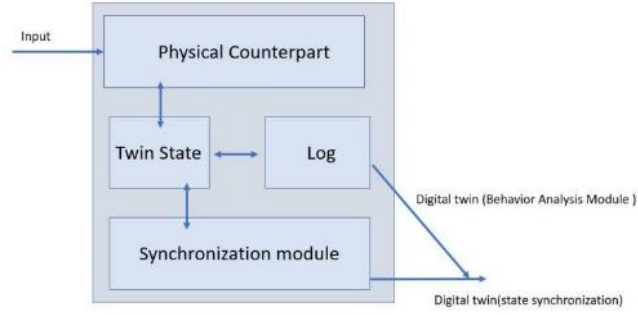

Fig. 3 The physical twin component

*3.5. State machine replication design*

From the previously research [11]-[13], we learn that there are many approaches that can combine state machines with the digital twin system. In our proposed state replication model, we predefine the key states of devices in each physical domain and virtual domain. Additionally, we transfer the key states at a regular time slot. In this case, this framework can be more generalized and can be used in any CPSs. In other words, we do not need to consider too much about the structure of digital twin systems and the synchronization frequencies.

There are many different approaches for state machines in digital twin systems. Because the physical twin is the main controller of the whole system, we could denote $f_d : S_d \times S_{d'} \to S_d$ and $g_d : S_d \to S_{d'}$ to represent the synchronization. Furthermore, from the previously section, we understand that the virtual domain is driven by the physical domain. Hence, we assume the states of physical twins can be divided into three parts and each part is independently. Then, we let:

$$S_d = S_{1d} \cup S_{2d} \cup S_{3d} \quad (1)$$
$$S_{1d} \cap S_{2d} = S_{1d} \cap S_{3d} = S_{2d} \cap S_{3d} = \varnothing \quad (2)$$

Similarly, the states of the digital twins can be divided into two parts and each part is independently. We can let:

$$S_{d'} = S_{1d'} \cup S_{2d'} \quad (3)$$
$$S_{1d} \cap S_{2d} = \varnothing \quad (4)$$

The following Fig. 4 explains the synchronization process. Before the first synchronization, we assume the key state of a physical twin as $S_{1d}$. In this time, the operator sends an input $i_d$ to the physical twin. Hence, there is no exact same state in the digital twin at this time. After the synchronization, the first key state of the digital twin receives the data, which is the input $i_d$ based on the state $S_{1d}$. At next time slot, which is t+1, The first key state of the digital twin should be equal to the first key state of physical twin. This achieves the Requirement 1 from section 3C. Then, we assume the physical twin receives another input $i_d$. Because the physical domain has the records of synchronized states of the digital twin, it could calculate the differences $\triangle S_d$ and send this data to the virtual environment. The digital twin uses this $\triangle S_d$ to move to the next key state. Furthermore, at the next time slot t+2, the key states of both twins are consistent. This achieves Requirement 2 from section 3C. In our framework, the operator can also send the inputs to the digital twin. we assume the digital twin receives an input $i_{d'}$ at the time t+1. From the virtual environment perspective, the digital twin would move to the next key state $s_{2d'}$ by executing the input $i_{d'}$. Moreover, the digital twin has the records of synchronized key states of the physical twin as well. It then calculates the $\triangle S_{d'}$ and send to the physical domain. The physical twin can use this data to construct the next stage. This achieves Requirement 3 from section 3C. Hence, at the time t+2, if there are no inputs, the key states of the physical twin and digital twin could keep the same. Finally, we could get the result:

$$S_{2d} = S_{1d'} \tag{5}$$
$$S_{3d} = S_{2d} \tag{6}$$

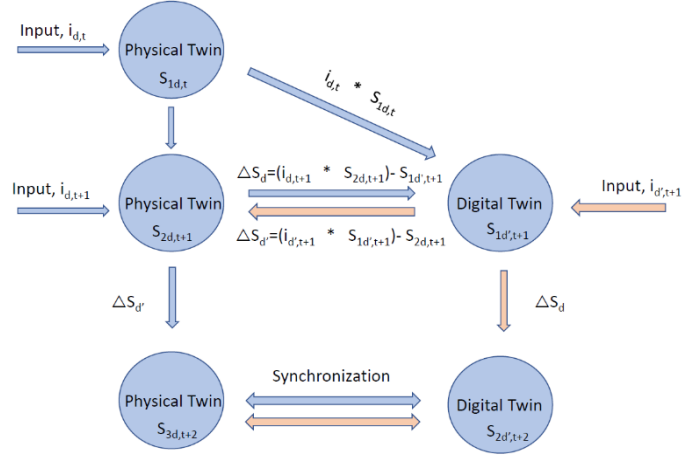

Fig. 4 The state machine framework

## 4. Conclusion

Inspired from security concerns in CPSs, we introduced a digital twin framework for the CPSs. The main purpose of our contribution is to provide secure synchronization. Unlike other approaches, our proposed system has predefined some key states of each device and transfer those key states based on a regular time slot. By only transfer the key states, this framework solves the problem of latency synchronization between the physical domain and the virtual domain. Furthermore, we demonstrated the proposed synchronization meet the requirements of this project. In the future, we need to consider how to predefine the key state of each device in an optimal way.